\documentclass[a4paper]{article}

\usepackage{INTERSPEECH2019}
\usepackage{setspace}
\usepackage{multirow}
\usepackage{color}
\usepackage{subfigure}
\usepackage{footnote}
\title{Hierarchical Pooling Structure for Weakly Labeled Sound Event Detection}
\name{Ke-Xin He*, Yu-Han Shen*, Wei-Qiang Zhang\thanks{The first two authors contributed equally.}\thanks{The corresponding author is Wei-Qiang Zhang.}\thanks{This work was supported by the National Natural Science Foundation of China under Grant No. U1836219.}}
\address{
 Department of Electronic Engineering, Tsinghua University, Beijing 100084, China}
\email{hekexinchn@163.com, yhshen@hotmail.com, wqzhang@tsinghua.edu.cn}

\begin{document}
\maketitle
\begin{abstract}
Sound event detection with weakly labeled data is considered as a problem of multi-instance learning. And the choice of pooling function is the key to solving this problem. In this paper, we proposed a hierarchical pooling structure to improve the performance of weakly labeled sound event detection system. Proposed pooling structure has made remarkable improvements on three types of pooling function without adding any parameters. Moreover, our system has achieved competitive performance on Task 4 of Detection and Classification of Acoustic Scenes and Events  (DCASE) 2017 Challenge using hierarchical pooling structure.

\end{abstract}

\section{Introduction}
The aim of sound event detection (SED) is to detect what types of sound events occur in an audio stream and furthermore, locate the onset and offset times of sound events.

Traditional approaches of SED depend on strongly labeled data, which provides the type and its timestamp (onset and offset time) of each sound event occurrence. But such annotation is too consuming to acquire. In consequence, many researchers begin to focus on the detection of sound events using weakly labeled training data. \emph{Weak label} represents that training data are annotated with only the presence of sound events and no timestamps are provided.

 Google released the weakly labeled Audio Set \cite{audioset} in 2017, which boosted the development of relevant research community. The Detection and Classification of Acoustic Scenes and Events (DCASE) 2017 Challenge launched a task of large-scale weakly supervised sound event detection for smart cars \cite{task}, and it employed a subset of Audio Set.

Common solutions to SED with weak label are based on Multi-Instance Learning (MIL). In MIL, the groundtruth label of each instance is unknown. Instead, we only know the groundtruth label of bags, each containing many instances. A bag is labeled negative if all instances are negative; a bag is labeled positive if at least one instance in it is positive. As shown in Figure 1, in MIL for SED, an audio clip can be considered as a bag, each consisting of several frames. For a specific class of sound events, a clip is labeled positive if target sound event occurs in at least one frame.

\begin{figure}[tb]%{1.0\linewidth}
\centering{\includegraphics[width=0.85\linewidth]{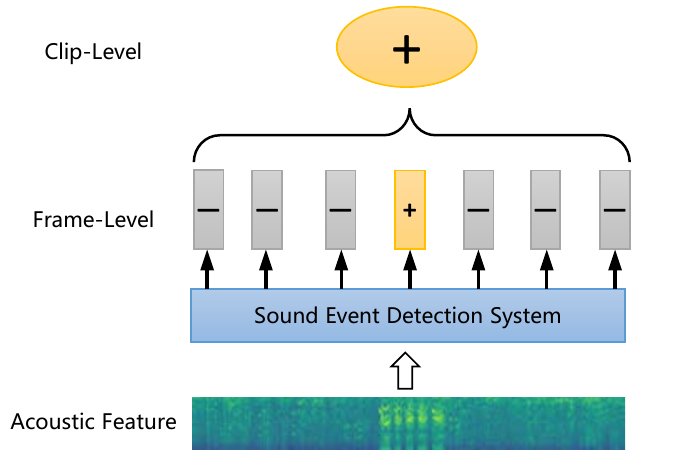}}
\caption{Illustration of Multi-Instance Learning System for sound event detection with weakly labeled data.}
\label{fig:fig1}
\vspace{-0.4cm}
\end{figure}

To solve the problem of MIL for SED, we usually use neural networks to predict the probabilities of each sound event class occurring in each frame. Then, we need to aggregate the frame-level probabilities into a clip-level probability for each class of sound events. Standard approaches to aggregating the probabilities include max-pooling and average-pooling, and there are also many variants and developments. Kong et al. \cite{Kong} proposed an attention model as pooling function, which has been adopted in many works \cite{xu2018large,jiakai2018mean,kong2019sound}. McFee et al. \cite{McFee} proposed a family of adaptive pooling operators. Wang et al. \cite{Wang} compared five pooling functions for SED with weak labeling.

In our paper, we proposed a hierarchical pooling structure to give a better supervision for neural network learning. Proposed pooling structure has improved the performance of three types of pooling functions without any added parameters. We evaluate our methods on DCASE 2017 Challenge Task 4 and our model has shown excellent performance.

%The rest of this paper is organized as follows: we introduce our methods in detail, mainly including baseline, pooling function and hierarchical pooling structure in Section 2. Relevant information about our experiments are illustrated in Section 3. We present the experimental results and related analysis in Section 4. Finally, we draw our conclusion in Section 5.

\section{Methods}
\label{sec:methods}
\subsection{Baseline System}
\label{ssec:baseline}
The mainstream Convolutional Recurrent Neural Network (CRNN) system is implemented as our baseline. The overview of baseline system is illustrated in Figure 2.

\begin{figure*}[tb]
\centerline{\includegraphics[width=0.9\linewidth]{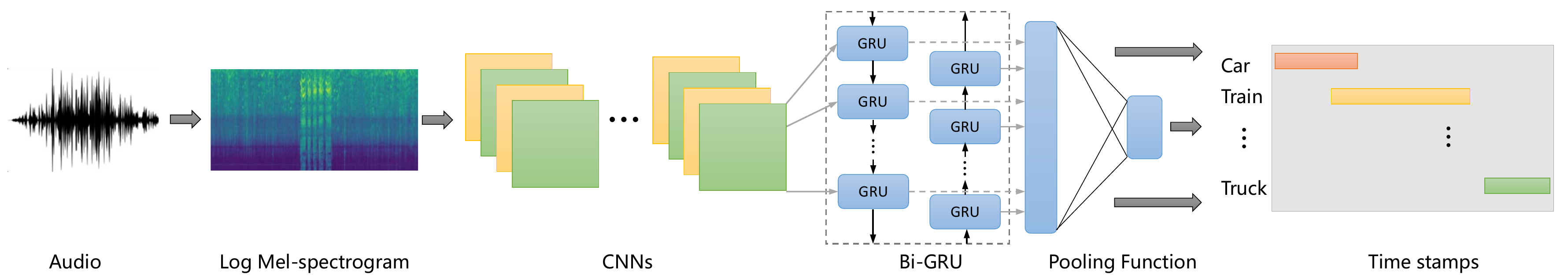}}
\caption{Overview of baseline system.}
\label{fig:baseline}
\end{figure*}

We use log mel spectrogram as acoustic feature. The input feature will pass through several Convolutional layers, a Bi-directional Gated Recurrent Unit (Bi-GRU) and a dense layer with sigmoid activation to produce predictions for frame-level presence probabilities of each sound event class.

The architecture of neural networks in our work is similar to that in \cite{xu2018large}. As shown in Figure 3, the Convolutional Neural Network (CNN) part consists of four convolutional blocks and a single convolutional layer. Each block contains two gated convolutional layers \cite{glu}, batch normalization \cite{BN}, dropout \cite{dropout} and a max-pooling layer. Max-pooling layers are adopted on both time axis and frequency axis. Note that the frame rate has reduced from 50 Hz to 12.5 Hz due to the max-pooling operations on time axis. The extracted features over different convolutional channels are stacked to the frequency axis before being fed into the Recurrent Neural Network (RNN) part.

\begin{figure}[tb]
\centerline{\includegraphics[width=0.8\linewidth]{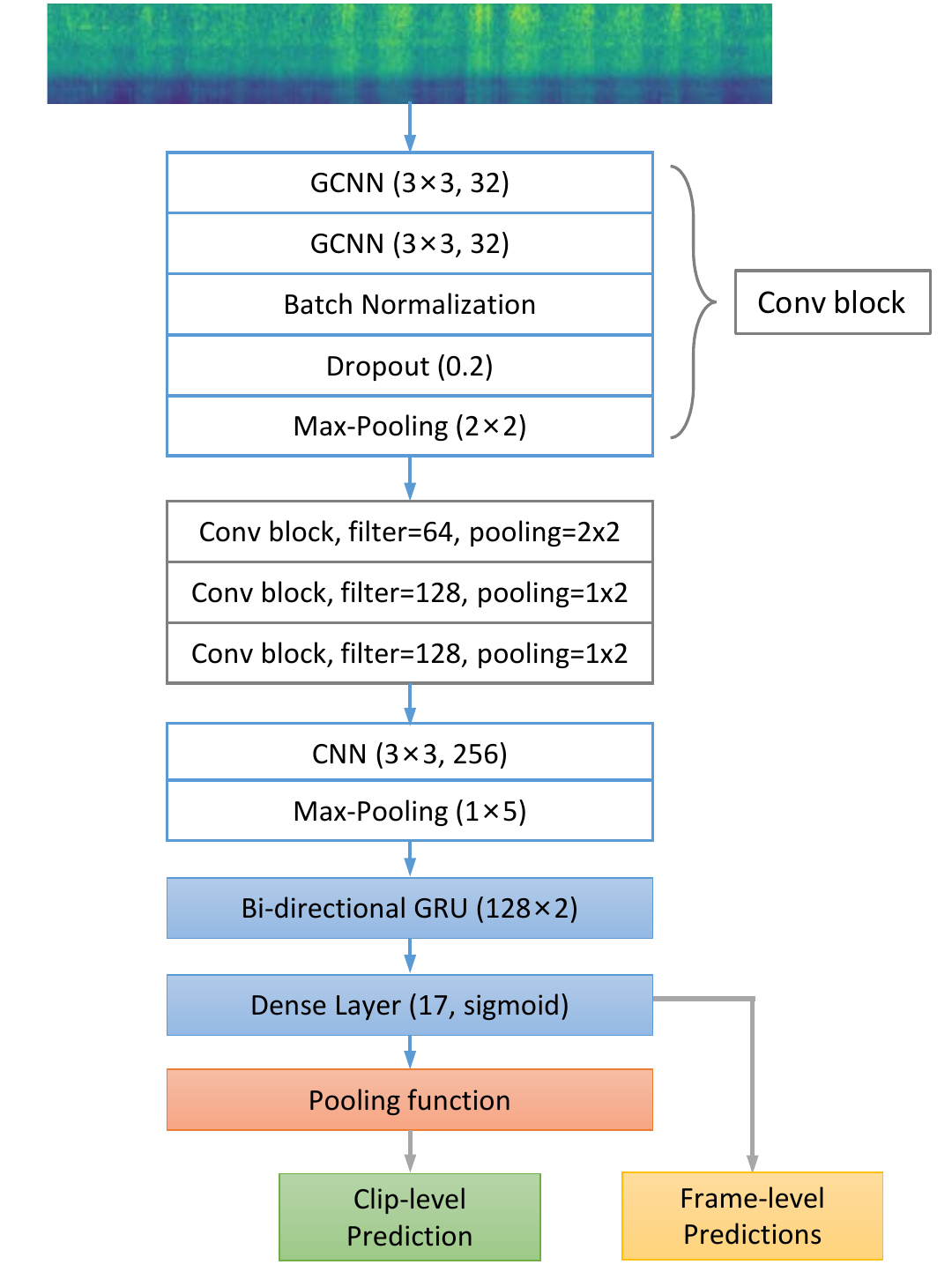}}
\caption{Architecture of neural networks. The first and second dimensions of convolutional kernels and strides represent the time axis and frequency axis respectively. The size of all convolutional kernels is $3\times 3$.}
\vspace{-0.4cm}
\end{figure}

The RNN part in our work is based on Bi-GRU. The outputs of forward and backward GRU are concatenated to get final outputs. The hyper-parameters are included in Figure 3.

Finally, a pooling function is adopted to calculate the presence probability of each sound event class in a 10-second audio clip. The choice and usage of pooling function will be specifically explained in the following parts of this section.

For testing, in order to locate the detected sound events, a threshold $\theta$ is set to the frame-level predictions. Then, we use post-processing methods including median filter and ignoring noise to get the onset and offset times of detected events.

\subsection{Pooling function}
\label{ssec:pooling}
As mentioned above, the design of pooling function is an essential issue in weakly labeled sound event detection. Wang et al. \cite{Wang} made a comprehensive comparison of five pooling functions (max pooling, average pooling, linear softmax, exponential softmax and attention) in MIL for SED. Those pooling functions are introduced as follows.

\begin{table}[thb]
  \centering
  \caption{Definition of five pooling functions}
  \renewcommand{\arraystretch}{2.7}
  \scalebox{0.82}
  {\begin{tabular}{ccc}

    \toprule
     \specialrule{0em}{-3pt}{-3pt}
     \textbf{Pooling function} & \textbf{Definition} & \textbf{Weight value} \\
     \specialrule{0em}{-1pt}{-1pt}
    \midrule
    \specialrule{0em}{-1.5pt}{-1.5pt}
     \textbf{Max pooling} & $y = \max\limits_{i}{x_i}$ &  $$$\displaystyle  w_i=\begin{cases}
                                                            1,& \text{$i = \mathop{\arg\max}\limits_{i}{x_i}$   }\\
                                                            0,& \text{else}
                                                            \end{cases}$$$ \\

    \specialrule{0em}{-3pt}{-3pt}
    % \textbf{Average pooling} & $y = \frac{1}{n}\sum{x_i}$ & $w_i = \frac{1}{n}$\\
    \textbf{Average pooling} & $\displaystyle y = \frac{1}{n}\sum{x_i}$ & $\displaystyle w_i = \frac{1}{n}$\\
    \specialrule{0em}{-2pt}{-2pt}
   \textbf{Linear softmax} & $\displaystyle y = \frac{\sum{{x_i}^2}}{\sum{x_i}}$  & $w_i = x_i$  \\
     \textbf{Exp. softmax}   &   $\displaystyle y = \frac{\sum{{x_i}{\exp{(x_i)}}}}{\sum{\exp{(x_i)}}}$ & $w_i = \exp{(x_i)}$ \\

     \textbf{Attention} & $\displaystyle y =  \frac{\sum{w_{i}x_{i}}}{\sum{w_i}}$ & $w_i = h(\textbf{u})$\\
    \bottomrule
    \end{tabular}}
\vspace{-0.5cm}
\end{table}

Let $x_i \in [0, 1]$ be the predicted probability of a specific event class occurring at the $i$-th frame. We need a pooling function to make a clip-level prediction. Let $y \in [0, 1]$ be the clip-level probability, and we have the following equation:
\begin{spacing}{0.78}
\begin{equation}
   %y = \frac{\sum{w_{i}x_{i}}}{\sum{w_i}}
   y = \frac{\sum_{i=1}^{N}{w_{i}x_{i}}}{\sum_{i=1}^{N}{w_i}}
\end{equation}
\end{spacing}
\noindent{where $w_i$ is the weight coefficient for $x_i$, and $N$ is the number of frames in a clip. Shown in Table 1 is the formula to calculate the weight values $w_i$ for five types of pooling functions.}

In the case of attention pooling function, the weight value $w_i$ is learnt by a dense layer with softmax activation. And its input \textbf{u} is the same as the input of the dense layer producing $x_i$. It is obvious from Table 1 that $w_i$ is a function of $x_i$ or \textbf{u}, so we denote this function as:
\begin{spacing}{0.65}
\begin{equation}
   w_i = f(x_i; \textbf{u})
\end{equation}
\end{spacing}

\subsection{Hierarchical pooling structure}
\label{ssec:Hierarchical}
\begin{figure*}[t]
\centerline{\includegraphics[width=0.9\linewidth]{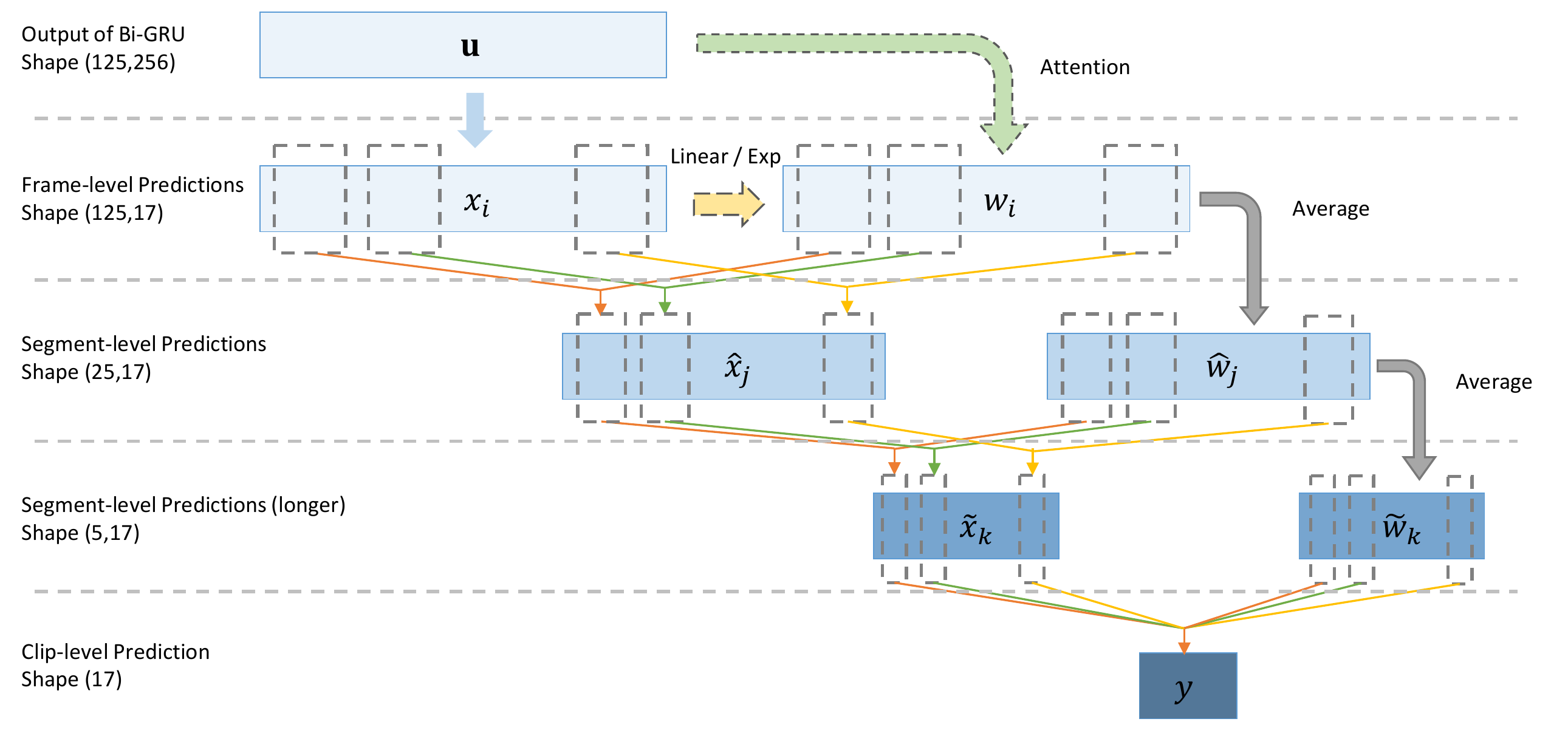}}
\caption{Three-stage hierarchical pooling structure. In linear and exponential softmax pooling, the frame-level weights $w_i$ derive from frame-level predictions $x_i$ ; in attention pooling, they are learnt from the output of Bi-GRU. In the first stage, every five frames are aggregated together to get segment-level predictions $\hat{x}_{j}$; the weights of every five frames are averaged to get segment-level weights $\hat{w}_{j}$. In the second stage, every five segments are aggregated to get longer-segment-level predictions $\widetilde{x}_k$ and every five segment-level weights are averaged to get longer-segment-level weights  $\widetilde{w}_k$. In the end, $\widetilde{x}_k$ and $\widetilde{w}_k$ are aggregated to get final clip-level prediction.}
\end{figure*}

Instead of aggregating all $N$ frame-level predictions $x_i$ to a clip-level prediction $y$ at once, we firstly group $N$ frames into several segments with the length of $M$ to make segment-level predictions $\hat{x}_{j}$. At the same time, the weight values $w_i$ are also weighted averaged using themselves as weights  to obtain segment-level weights $\hat{w}_{j}$. Finally, we use the segment-level predictions $\hat{x}_{j}$ and weights $\hat{w}_{j}$ to get the clip-level prediction $y$. The entire process is illustrated by the following formulas.
\begin{spacing}{0.75}
\begin{equation}
   \hat{x}_{j} = \frac{\sum_{i=1+(j-1)M}^{jM}{w_{i}x_{i}}}{\sum_{i=1+(j-1)M}^{jM}{w_i}}, j = 1, 2, ..., N/M
\end{equation}

\begin{equation}
  \begin{split}
   \hat{w}_{j} &= \frac{\sum_{i=1+(j-1)M}^{jM}{w_{i}}^{2}}{\sum_{i=1+(j-1)M}^{jM}{w_{i}}}, j = 1, 2, ..., N/M
   \end{split}
\end{equation}

\begin{equation}
   y = \frac{\sum_{j=1}^{N/M}{\hat{w}_{j}\hat{x}_{j}}}{\sum_{j=1}^{N/M}{\hat{w}_{j}}}
\end{equation}
\end{spacing}

\subsection{Analysis of hierarchical pooling structure}
Before we discuss this structure in depth, we would like to arrive at a proposition: \emph{the accuracy of $\hat{x}_{j}$ is larger than that of $x_i$ in a well-trained system}. This proposition is intuitively reasonable because it is easier for the system to output correct predictions when the required time resolution gets longer.

According to the theoretical discussion in \cite{Wang}, the process of weight updating is related to $\frac{\partial{y}}{\partial{x_i}}$ and $\frac{\partial{y}}{\partial{w_i}}$. We take linear softmax pooling function as an example to interpret the function of proposed pooling structure.

In the case of normal single pooling structure, $w_i = x_i$,
\begin{spacing}{0.8}
\begin{equation}
   \frac{\partial{y}}{\partial{x_i}} = \frac{2x_i - y}{\sum_{k=1}^{N}{x_k}}
\end{equation}
\end{spacing}

In the case of hierarchical pooling structure,
\begin{spacing}{0.8}
\begin{equation}
    \begin{split}
   \hat{w}_{j} &= \frac{\sum_{i=1+(j-1)M}^{jM}{w_{i}}^{2}}{\sum_{i=1+(j-1)M}^{jM}{w_{i}}} \\
               &= \frac{\sum_{i=1+(j-1)M}^{jM}{x_{i}}^{2}}{\sum_{i=1+(j-1)M}^{jM}{x_{i}}}
   \end{split}
\end{equation}
\begin{equation}
  \begin{split}
   \frac{\partial{y}}{\partial{x_i}}  &= \sum_{l=1}^{N/M}
   {(\frac{\partial{y}}{\partial{\hat{x}_{l}}}{\frac{\partial{\hat{x}_{l}}}{\partial{x_i}}}
   +\frac{\partial{y}}{\partial{\hat{w}_{l}}}{\frac{\partial{\hat{w}_{l}}}{\partial{x_i}}})}\\
   &= \frac{x_i(4\hat x_j -2y) -2 {\hat x_j}^{2}+y \hat x_j}{\sum_{l=1}^{N/M}\hat x_l \sum_{n=1+(j-1)M}^{jM}x_i}, \quad  j=\lceil{\frac{i}{M}}\rceil
   \end{split}
\end{equation}
\end{spacing}
\vspace{0.1cm}

 As shown in Equation 6 and Equation 8, compared with single pooling structure, the segment-level prediction $\hat{x}_{j}$ also contributes to the update of frame-level prediction $x_i$ in hierarchical pooling structure. As segment-level prediction is more accurate than frame-level prediction, we believe proposed hierarchical pooling structure can provide a better supervision for neural network learning.

Detailed mathematical derivation and analysis of all five pooling functions are available in the appendix. We proved that proposed structure would make no difference on max and average pooling, so we conducted our experiments using the other three pooling functions.

The hierarchical pooling structure used in our work is illustrated in Figure 4. It is a three-stage pooling structure. The number of predicted probabilities for a certain class of sound events in an audio clip decreases from 125 to 25, and then 5, and finally 1.

\section{Experiments}
\label{sec:experiments}
\begin{table*}[t]
  \centering
  \caption{Performance of single and hierarchical pooling structure, in terms of ER (lower is better) and $F_{1}$-score (\%) (higher is better).}
  \renewcommand{\arraystretch}{1.2}
  \scalebox{0.83}
  {\begin{tabular}{c|c|cccc|ccc|cccc|ccc}
    \hline
    \multicolumn{2}{c}{} & \multicolumn{7}{|c|}{Single Pooling Structure} &  \multicolumn{7}{c}{Hierarchical Pooling Structure} \\

    \cline{3-16}
    \multicolumn{2}{c|}{} & Sub. & Del. & Ins. & ER & Pre. & Rec. & $F_{1}$ & Sub. & Del. & Ins. & ER & Pre. & Rec. & $F_{1}$  \\
    \hline

    \multirow{3}*{Development} & Linear & 0.25 & 0.18 & 0.36 & \textbf{0.79}& 39.00 & 47.01 & \textbf{42.63} & 0.19 & 0.40 & 0.17 & \textbf{0.76} \color{red}{(3.8\%$\downarrow$)}& 53.07 & 41.31 & \textbf{46.46}\ \ \color{red}{ (9.0\%$\uparrow$)}\\

    ~ & Exp. & 0.29 & 0.35 & 0.18 & \textbf{0.82}& 44.67 & 37.24 & \textbf{40.62}& 0.27 & 0.26 & 0.26 & \textbf{0.79} \color{red}{(3.7\%$\downarrow$)}& 45.90 & 45.72 & \textbf{45.81} \color{red}{(12.8\%$\uparrow$)} \\

    ~ & Att. & 0.30 & 0.34 & 0.19 & \textbf{0.83}& 44.68 & 36.97 &\textbf{ 40.46} & 0.25 & 0.33 & 0.21 & \textbf{0.79} \color{red}{(4.8\%$\downarrow$)}& 48.17& 42.51 & \textbf{45.16} \color{red}{(11.6\%$\uparrow$)} \\
    \hline
    \multirow{3}*{Evaluation} & Linear & 0.21 & 0.36 & 0.18 & \textbf{0.76}& 53.40 & 43.19 & \textbf{47.76} & 0.19 & 0.30 & 0.20 & \textbf{0.69} \color{red}{(9.2\%$\downarrow$)}& 56.39 & 50.78 & \textbf{53.44} \color{red}{ (11.8\%$\uparrow$)}\\

    ~ & Exp. & 0.23 & 0.35 & 0.23 & \textbf{0.81}& 48.35 & 43.78 & \textbf{45.95} & 0.23 & 0.28 & 0.22 & \textbf{0.73} \color{red}{(9.9\%$\downarrow$)}& 53.40 & 51.38 & \textbf{52.37} \color{red}{(14.0\%$\uparrow$)}\\

    ~ & Att. & 0.21 & 0.31 & 0.27 & \textbf{0.79}& 46.12 & 44.43 &\textbf{ 45.26 }& 0.21 & 0.28 & 0.24 & \textbf{0.73} \color{red}{(7.6\%$\downarrow$)}& 52.27 & 50.58 & \textbf{51.41} \color{red}{(13.6\%$\uparrow$)}\\
    \hline
    \end{tabular}}
\vspace{-0.1cm}
\end{table*}

\subsection{Dataset}
\label{ssec:dataset}
We demonstrated our experiments on task 4 of DCASE 2017 Challenge \cite{task}. This task contains 17 classes of sound events. The dataset is a subset of Audio Set \cite{audioset}. The training set has weak labels denoting the presence of a given sound event in the videos soundtrack and no timestamps are provided. For testing and evaluation, strong labels with timestamps are provided for the purpose of evaluating performance.

\subsection{Experimental Setup}
\label{ssec:setup}
To extract log mel spectrogram feature, each audio is divided into frames of 40 ms duration with 50\% overlapping. The input of our system is a $500\times 80$ matrix, where $500$ denotes the number of frames and $80$ is the number of mel-filter bins.

Our model is trained using Adam optimizer \cite{Adam}. The initial learning rate is 0.001. The mini batch size is 128. The loss function is categorical cross entropy based on clip-level labels. We use early stop strategy when the validation loss stops degrading for 10 epochs.

\begin{figure}[tb]
\centerline{\includegraphics[width=0.98\linewidth]{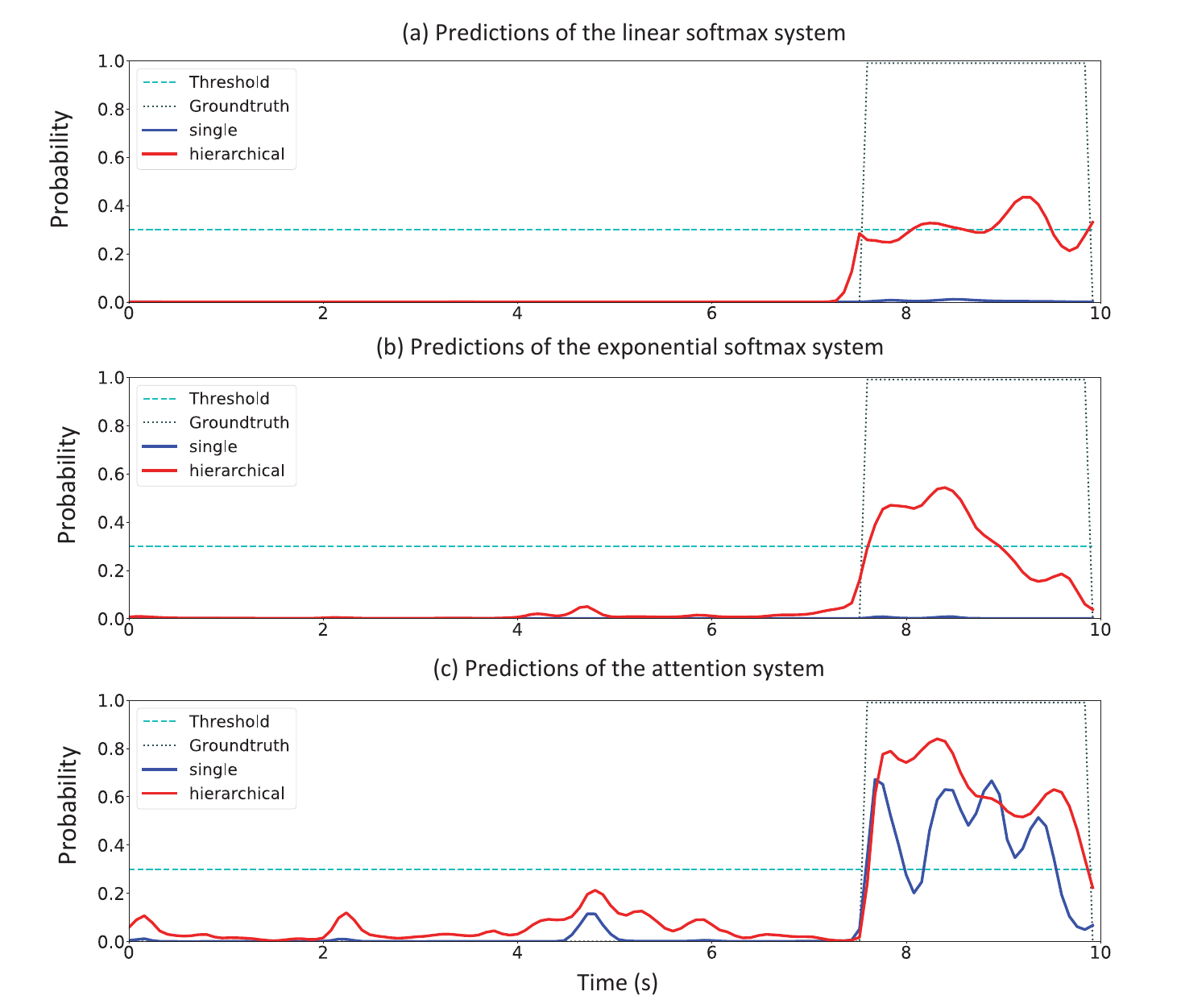}}
\caption{The frame-level predictions of three systems on an evaluation audio clip.}% named ' '.  The blue lines represent the frame-level predictions on single pooling structure while the red lines denote hierarchical pooling structure. The green dot-dash lines represent threshold and the black dot-dash lines represent groundtruth.}
\vspace{-0.4cm}
\end{figure}

\subsection{Metrics}
\label{ssec:Metrics}
According to the official instructions of DCASE 2017 Challenge \cite{task}, our method is evaluated based on two kinds of segment-based metrics: the primary metric is segment-based micro-averaged error rate (ER) and the secondary metric is segment-based micro-averaged $F_{1}$-score. ER is the sum of Substitution, Deletion and Insertion Errors, and $F_{1}$-score is the harmonic avergae of Precision and Recall. Each segment-based metric will be calculated in one-second segments over the entire test set. Detailed information can be found in \cite{task}. We use sed\_eval toolbox \cite{sed_eval} to compute the metrics.

\section{Results}
\label{sec:results}
\subsection{Experimental results}

\begin{table}[t]
  \centering
  \caption{Comparison with other methods, in terms of ER and $F_{1}$-score (\%). We compare proposed system with the following systems: (1)EMSI: 1st place in DCASE 2017; (2) Surrey: 2nd in DCASE 2017; (3) MLMS8: 3rd in DCASE 2017; (4) GCCaps: A Capsule Routing Network proposed in 2018; (5) Wang: Linear softmax system proposed in 2018. }
  %\renewcommand{\arraystretch}{1.2}
  %\scalebox{0.83}
  {\begin{tabular}{c|c|c|c|c}
    \hline
    ~ & \multicolumn{2}{|c|}{Development} &  \multicolumn{2}{c}{Evaluation} \\

    \cline{2-5}
    ~ & ER  & $F_{1}$ & ER  & $F_{1}$  \\
    \hline
    EMSI$^*$ \cite{lee2017ensemble}  & 0.71 & 47.1 & 0.66 & 55.5\\
    Surrey$^*$ \cite{xu2017surrey} & 0.72 & 49.7 & 0.73 & 51.8\\
    MLMS8$^*$ \cite{lee2017combining} & 0.84 & 34.2 & 0.75 & 47.1\\
    GCCaps \cite{iqbal2018capsule} & - - - & - - - & 0.76 & 46.3\\
    Wang \cite{Wang} & 0.79 & 45.4 & - - - & - - - \\
    \hline
    Proposed       & 0.76 & 46.5 & 0.69 & 53.4\\
    \hline
    \end{tabular}}
    \footnotesize{\\$^*$ system using model ensemble;\\
     - - - results not presented in paper.}
\vspace{-0.4cm}
\end{table}

We apply single pooling structure and proposed hierarchical pooling structure to three types of pooling functions. The performance on development and evaluation dataset is shown in Table 2. The percentage in red represents the change rate from single pooling structure to hierarchical pooling structure. Proposed structure can make remarkable improvements in all situations without adding any parameters. It is safe to draw a conclusion that hierarchical pooling structure can improve the performance of weakly-labeled sound event detection system significantly.
Besides, linear softmax pooling function outperforms the other pooling functions in all conditions, which corresponds with the experimental results in \cite{Wang}.

Figure 5 illustrates the frame-level predictions of single and hierarchical pooling structures on three pooling functions. In this audio, the sound of train occurs from 7.574 s to 10 s. In linear and exponential softmax, single pooling structure cannot output any positive predictions; on the contrary, hierarchical pooling structure can correctly detect target event. In attention pooling, the predicted probabilities of hierarchical structure are also higher than single structure where the event occurs. Besides, the linear and exponential softmax are more likely to produce deletion errors while attention will result in more insertion errors. This also complies with the analysis in \cite{Wang}.

\subsection{Comparison with other methods}
Compared with other methods, the performance of our system is also competitive. We compare proposed system with the top 3 teams in DCASE 2017 Challenge and two methods proposed in 2018. Proposed system can outperform most methods except the top 1 system in DCASE 2017 Challenge \cite{lee2017ensemble}. Note that the top 1 team utilized the ensemble of multiple systems, which significantly improved its performance. Our system can achieve comparable performance without ensemble.

\section{Conclusion}
In this paper, we proposed a hierarchical pooling structure to solve the problem of Multi-Instance Learning. We applied this strategy to develop a weakly-labeled sound event detection system. Our proposed method can effectively improve the performance in three types of pooling functions without adding any parameters. Besides, our best system can achieve comparable performance with the state-of-the-art systems without the techniques of ensemble. We believe our method can be applied in more applications of Multi-Instance Learning in addition to the field of weakly labeled sound event detection.

\bibliographystyle{IEEEtran}
\bibliography{hierarchical}

\appendix
\include{appendixA} \label{appendixprocedure}
\newpage{\pagestyle{empty}\cleardoublepage}
\renewcommand\thetable{\Alph{section}-\arabic{table}}
\renewcommand\theequation{A-\arabic{equation}}
\setcounter{table}{0}
\setcounter{equation}{0}
\section{Erratum}
\textbf{Comment: We figure out some errors in our paper, which has been published in the proceedings of Interspeech 2019. In order to correct the errors, we update the Arxiv version. If any of you is interested in our work, please refer to the lastest version on Arxiv. If you have any further questions, please feel free to contact the authors.}

The main error in our paper is the formula of segment-level weights $w_j$ in hierarchical pooling structure, i.e. Equation (4) in the body of this paper. 

The original formula is
\begin{equation}
  \begin{split}
   \hat{w}_{j} &= \frac{\sum_{i=1+(j-1)M}^{jM}{w_{i}}}{M}, j = 1, 2, ..., N/M
   \end{split}
\end{equation}

The corrected formula is
\begin{equation}
  \begin{split}
   \hat{w}_{j} &= \frac{\sum_{i=1+(j-1)M}^{jM}{w_{i}}^{2}}{\sum_{i=1+(j-1)M}^{jM}{w_{i}}}, j = 1, 2, ..., N/M
   \end{split}
\end{equation}

 Our motivation is that segment-level prediction is more accurate than frame-level prediction and it is easier to get correct predictions when the required time resolution gets longer. So we let the groundtruth clip-level labels supervise the training of small segment-level predictions and get an accurate segment-level prediction first, instead of directly supervising the training of each frame. 

 Besides, there are many other methods to get $w_j$ in hierarchical pooling structure. For example, we can add two extra dense layers after Bi-GRU to get $\hat{w}_{j}$ and $\widetilde{w}_k$ in Figure 4. It can also achieve similar effects but requires a small number of additional parameters. 

We also did some experiments based on the wrong formula in the paper and the average of ER is similar to single pooling structure. But during experiments, we find that system performances have big fluctuation. For example, we use attention pooling function for five experiments and the ER on evaluation dataset is 0.80, 0.85, 0.79, 0.78, 0.80 respectively. Meanwhile, in order to locate the detected sound events, a threshold is set to the frame-level predictions. And we use post-processing methods including median filter and ignoring noise to get the onset and offset times of detected events. The evaluation performance is also sensitive to the parameter of threshold and post-processing. We think our system may meet with overfitting. In future work, we will evaluate whether our proposed method is general and robust on larger datasets.

\section{Appendix}

Detailed mathematical derivation and analysis of all five pooling functions are available in the appendix.

The loss function we use is cross-entropy loss:
\begin{equation}
L = -t \log{y} - (1-t) \log{(1-y)}
\end{equation}
where $t = 0$ or $1$, is the groundtruth label for a specific sound event in an audio clip, and $y \in [0, 1]$ is the predicted clip-level probability for the same event.

We decompose the gradient of $L$ with respect to the frame-level predictions $x_i$ and the frame-level weights $w_i$ using chain rule:

\begin{equation}
\frac{\partial L}{\partial x_{i}} = \frac{\partial L}{\partial y} \frac{\partial y}{\partial x_{i}},
\frac{\partial L}{\partial w_{i}} = \frac{\partial L}{\partial y} \frac{\partial y}{\partial w_{i}}
\end{equation}

Considering the term $\displaystyle \frac{\partial L}{\partial y}$, we have:
\begin{equation}
\begin{split}
\frac{\partial L}{\partial y} &= -\frac{t}{y}  + \frac{1-t}{1-y}\\
&= \begin{cases}
         \frac{1}{1-y},& \text{$t=0$}\\
         -\frac{1}{y},& \text{$t=1$}
   \end{cases}
\end{split}
\end{equation}

It is obvious that this term is decided by the label $t$, so we focus on $\displaystyle  \frac{\partial y}{\partial x_{i}}$ and
$\displaystyle \frac{\partial y}{\partial w_{i}}$ in the following discussions.

Before proceeding into the calculation process, let us review the expression of $y$ in our hierarchical pooling structure.

\begin{equation}
  \label{eqn:pooling_equation3}
   \hat{x}_{j} = \frac{\sum_{i=1+(j-1)M}^{jM}{w_{i}x_{i}}}{\sum_{i=1+(j-1)M}^{jM}{w_i}}, j = 1, 2, ..., M
\end{equation}

\begin{equation}
  \begin{split}
   \hat{w}_{j} &= \frac{\sum_{i=1+(j-1)M}^{jM}{w_{i}}^{2}}{\sum_{i=1+(j-1)M}^{jM}{w_{i}}}, j = 1, 2, ..., N/M
   \end{split}
\end{equation}

\begin{equation}
   y = \frac{\sum_{j=1}^{N/M}{\hat{w}_{j}\hat{x}_{j}}}{\sum_{j=1}^{N/M}{\hat{w}_{j}}}
\end{equation}

So $y$ is a weighted sum of $\hat{x}_j$ with weights $\hat{w}_j$.

\begin{equation}
    \begin{split}
\frac{\partial y}{\partial x_{i}} &= \sum_{l=1}^{N/M}(\frac{\partial y}{\partial \hat{x}_l}\frac{\partial \hat{x}_l} {\partial x_{i}}+ \frac{\partial{y}}{\partial{\hat{w}_{l}}}{\frac{\partial{\hat{w}_{l}}}{\partial{x_i}}}) \\
&= \frac{\partial y}{\partial \hat{x}_j}\frac{\partial \hat{x}_j}{\partial x_{i}} + \frac{\partial y}{\partial \hat{w}_j}\frac{\partial \hat{w}_j}{\partial x_{i}}, j = \lceil{\frac{i}{M}}\rceil
   \end{split}
\end{equation}

The four components are calculated as follows:
\begin{equation}
\frac{\partial y}{\partial \hat x_{j}} = \frac{\hat w_j}{\sum_{l=1}^{N/M} \hat w_l}
\end{equation}

\begin{equation}
    \begin{split}
\frac{\partial \hat x_j}{\partial x_i} &= \frac{\mathrm{d}\hat x_j}{\mathrm{d} x_i} + \frac{\partial \hat x_j}{\partial w_i} \frac{\partial w_i}{\partial x_i} \\
&=\frac{w_i}{\sum_{n=1+(j-1)M}^{jM} w_n} + \frac{x_i - \hat x_j}{\sum_{n=1+(j-1)M}^{jM} w_n} \frac{\partial w_i}{\partial x_i}
   \end{split}
\end{equation}

\begin{equation}
\frac{\partial y}{\partial \hat w_{j}} = \frac{\hat x_j - y}{\sum_{l=1}^{N/M} \hat w_l}
\end{equation}

\begin{equation}
    \begin{split}
\frac{\partial \hat w_j}{\partial x_i} &= \frac{\partial \hat w_j}{\partial w_i} \frac{\partial w_i}{\partial x_i} \\
&=\frac{2w_i - \hat w_j}{\sum_{n=1+(j-1)M}^{jM}w_n} \frac{\partial w_i}{\partial x_i}
   \end{split}
\end{equation}

Here, $\displaystyle \frac{\partial w_i}{\partial x_{i}}$ relies on the choice of pooling functions.

Hence we summarize as follows:
%\begin{footnotesize}
\begin{equation}
    \begin{split}
    \begin{aligned}
\frac{\partial y}{\partial x_i} &= \frac{\hat w_j}{\sum_{l=1}^{N/M}\hat w_l} \left ( \frac{w_i}{\sum_{n=1+(j-1)M}^{jM} w_n} \right. 
  \\& \; \; \; \; \; \left.+ \frac{x_i - \hat x_j}{\sum_{n=1+(j-1)M}^{jM} w_n} \frac{\partial w_i}{\partial x_i} \right ) \\
& \; \; \; \; \;  + \frac{\hat x_j -y}{\sum_{l=1}^{N/M}\hat w_l} \left ( \frac{2w_i - \hat w_j}{\sum_{n=1+(j-1)M}^{jM} w_n} \frac{\partial w_i}{\partial x_i}\right ) \\
&= \frac{ \hat w_j w_i + \left [ \left ( x_i - \hat x_j \right ) \hat w_j + \left ( \hat x_j - y \right ) \left ( 2w_i - \hat w_j \right )\right ] \frac{\partial w_i}{\partial x_i} }{\sum_{l=1}^{N/M}\hat w_l \sum_{n=1+(j-1)M}^{jM}w_n}
   \end{aligned}
   \end{split}
\end{equation}
%\end{footnotesize}

In the case of average pooling function, $\displaystyle w_i = \frac{1}{N}$,
\begin{equation}
    \begin{split}
\frac{\partial w_i}{\partial x_{i}} = 0
   \end{split}
\end{equation}

\begin{equation}
\frac{\partial y}{\partial x_i} = \frac{\hat w_j w_i}{\sum_{l=1}^{N/M} \hat w_l \sum_{n=1+(j-1)M}^{jM}w_n}=\frac{1}{N}
\end{equation}

In the case of max pooling function,
\begin{equation}
w_i=\begin{cases}
1,& \text{$i = \mathop{\arg\max}\limits_{i}{x_i}$   }\\
0,& \text{else}
\end{cases}
\end{equation}
so we have:

\begin{equation}
\frac{\partial w_i}{\partial x_{i}} = 0
\end{equation}

\begin{equation}
\frac{\partial y}{\partial x_{i}} = \begin{cases}
1,& \text{$i = \mathop{\arg\max}\limits_{i}{x_i}$   }\\
0,& \text{else}
\end{cases}
\end{equation}

In the case of linear softmax pooling function, $w_i = x_i$,
\begin{equation}
    \begin{split}
\frac{\partial w_n}{\partial x_{i}} = \begin{cases}
                                       1,& \text{$n = i$}\\
                                       0,& \text{else}
                                       \end{cases}
   \end{split}
\end{equation}

\begin{equation}
    \begin{split}
\frac{\partial y}{\partial x_i} &= \frac{ \hat w_j w_i + (x_i - \hat x_j) \hat w_j + (\hat x_j - y)(2w_i - \hat w_j) }{\sum_{l=1}^{N/M} \hat w_l \sum_{n=1+(j-1)M}^{jM}w_n}  \\
&=\frac{x_i(4\hat x_j -2y) -2 {\hat x_j}^{2}+y \hat x_j}{\sum_{l=1}^{N/M}\hat x_l \sum_{n=1+(j-1)M}^{jM}x_i}
    \end{split}
\end{equation}

In the case of exponential softmax pooling function, $w_i = \exp{(x_i)}$,
\begin{equation}
    \begin{split}
\frac{\partial w_n}{\partial x_{i}} = \begin{cases}
                                       \exp{(x_i)},& \text{$n = i$}\\
                                       0,& \text{else}
                                       \end{cases}
   \end{split}
\end{equation}

\begin{equation}
    \begin{split}
\frac{\partial y}{\partial x_i} &= \frac{ \hat w_j w_i +\left [  (x_i - \hat x_j) \hat w_j + (\hat x_j - y)(2w_i - \hat w_j) \right ] \exp (x_i) }{\sum_{l=1}^{N/M} \hat w_l \sum_{n=1+(j-1)M}^{jM}w_n}  \\
&=\frac{\left [ \hat w_j(1+x_i-2\hat x_j + y) + 2\exp (x_i)(\hat x_j - y) \right ] \exp (x_i)}{\sum_{l=1}^{N/M}\hat x_l \sum_{n=1+(j-1)M}^{jM}\exp (x_n)}
    \end{split}
\end{equation}

In the case of attention pooling function, $w_i$ is decided by the input of the last dense layer \textbf{u} instead of $x_i$,

\begin{equation}
\frac{\partial w_n}{\partial x_{i}} = 0
\end{equation}

\begin{equation}
\frac{\partial y}{\partial x_i} = \frac{\hat w_j w_i}{\sum_{l=1}^{N/M} \hat w_l\sum_{n=1+(j-1)M}^{jM} w_n}
\end{equation}

In this case, we should consider the item $\displaystyle \frac{\partial y}{\partial w_{i}}$  as well. The item is calculated as follows:

\begin{equation}
    \begin{split}
\frac{\partial y}{\partial w_i} &= \sum_{l=1}^{N/M} \frac{\partial y}{\partial \hat w_l} \frac{\partial \hat w_l}{\partial  w_i} + \frac{\partial y}{\partial \hat x_l} \frac{\partial \hat x_l}{\partial  w_i} \\
&=\frac{\partial y}{\partial \hat w_j} \frac{\partial \hat w_j}{\partial  w_i} + \frac{\partial y}{\partial \hat x_j} \frac{\partial \hat x_j}{\partial  w_i} \\
&=\frac{\hat x_j - y}{\sum_{l=1}^{N/M}\hat w_l}\frac{2w_i - \hat w_j}{\sum_{n=1+(j-1)M}^{jM}w_n} \\&\; \; \; \; \;+ \frac{\hat w_j}{\sum_{l=1}^{N/M}\hat w_l}\frac{x_i - \hat x_j}{\sum_{n=1+(j-1)M}^{jM}w_n} \\
&=\frac{2w_i(\hat x_j - y)+\hat w_j(x_i+y-2 \hat x_j)}{\sum_{l=1}^{N/M}\hat w_l \sum_{n=1+(j-1)M}^{jM}w_n}
    \end{split}
\end{equation}

The single pooling structure can be considered as a special case of hierarchical pooling structure in which $\hat{w}_j = w_i, \hat{x}_j = x_i $.

According to the analysis above, it is easy to notice that proposed hierarchical pooling structure will make no difference when applied to max pooling and average pooling functions. So we only analyze the other three pooling functions in our paper. As shown in above results, the segment-level prediction $\hat{x}_j$ will also contribute to weight updating during training. So we believe this kind of structure can give a better supervision for neural network learning.

\end{document}